\documentclass[aps,prl,twocolumn,superscriptaddress,letterpaper]{revtex4}

\usepackage{graphicx}% Include figure files
\usepackage{dcolumn}% Align table columns on decimal point
\usepackage{bm}% bold math
\usepackage{amsmath, amssymb}% 
\usepackage{epstopdf}
\usepackage{color}
\usepackage[titletoc,toc,title]{appendix}
\usepackage{lipsum}

\usepackage{hhline}
\usepackage{multirow}
\usepackage{float}
\floatstyle{plaintop}
\restylefloat{table}
\usepackage[]{changes} % final
\begin{document}
%\linenumbers

%Title of paper
\title{First Observation of Large Missing-Momentum $(e,e'p)$ Cross-Section Scaling and the onset of Correlated-Pair Dominance in Nuclei}

\newcommand*{\MIT}{Massachusetts Institute of Technology, Cambridge, Massachusetts 02139, USA}
\newcommand*{\MITindex}{1}
\affiliation{\MIT}
\newcommand*{\TAU}{School of Physics and Astronomy, Tel Aviv University, Tel Aviv 69978, Israel}
\newcommand*{\TAUindex}{2}
\affiliation{\TAU}
\newcommand*{\GWU}{The George Washington University, Washington DC 20052, USA}
\newcommand*{\GWUindex}{3}
\affiliation{\GWU}
\newcommand*{\ODU}{Old Dominion University, Norfolk, Virginia 23529, USA}
\newcommand*{\ODUindex}{4}
\affiliation{\ODU}
\newcommand*{\JLab}{Thomas Jefferson National Accelerator Facility, Newport News, Virginia 23606, USA}
\newcommand*{\JLabindex}{5}
\affiliation{\JLab}
\newcommand*{\ANL}{Physics Division, Argonne National Laboratory, Lemont, Illinois 60439, USA}
\newcommand*{\ANLindex}{6}
\affiliation{\ANL}
\newcommand*{\FNAL}{Theoretical Physics Department, Fermi National Accelerator Laboratory\\ , P.O. Box 500, Batavia, Illinois 60510, USA}
\newcommand*{\FNALindex}{7}
\affiliation{\FNAL}

\newcommand*{\CSUDH}{California State University, Dominguez Hills, Carson, CA 90747}
\newcommand*{\CSUDHindex}{8}
\affiliation{\CSUDH}
\newcommand*{\CANISIUS}{Canisius College, Buffalo, NY}
\newcommand*{\CANISIUSindex}{9}
\affiliation{\CANISIUS}
\newcommand*{\CMU}{Carnegie Mellon University, Pittsburgh, Pennsylvania 15213}
\newcommand*{\CMUindex}{10}
\affiliation{\CMU}
\newcommand*{\CUA}{Catholic University of America, Washington, D.C. 20064}
\newcommand*{\CUAindex}{11}
\affiliation{\CUA}
\newcommand*{\SACLAY}{IRFU, CEA, Universit\'{e} Paris-Saclay, F-91191 Gif-sur-Yvette, France}
\newcommand*{\SACLAYindex}{12}
\affiliation{\SACLAY}
\newcommand*{\CNU}{Christopher Newport University, Newport News, Virginia 23606}
\newcommand*{\CNUindex}{13}
\affiliation{\CNU}
\newcommand*{\UCONN}{University of Connecticut, Storrs, Connecticut 06269}
\newcommand*{\UCONNindex}{14}
\affiliation{\UCONN}
\newcommand*{\DUKE}{Duke University, Durham, North Carolina 27708-0305}
\newcommand*{\DUKEindex}{15}
\affiliation{\DUKE}
\newcommand*{\DUQUESNE}{Duquesne University, 600 Forbes Avenue, Pittsburgh, PA 15282 }
\newcommand*{\DUQUESNEindex}{16}
\affiliation{\DUQUESNE}
\newcommand*{\FU}{Fairfield University, Fairfield CT 06824}
\newcommand*{\FUindex}{17}
\affiliation{\FU}
\newcommand*{\FERRARAU}{Universita' di Ferrara , 44121 Ferrara, Italy}
\newcommand*{\FERRARAUindex}{18}
\affiliation{\FERRARAU}
\newcommand*{\FIU}{Florida International University, Miami, Florida 33199}
\newcommand*{\FIUindex}{19}
\affiliation{\FIU}
\newcommand*{\FSU}{Florida State University, Tallahassee, Florida 32306}
\newcommand*{\FSUindex}{20}
\affiliation{\FSU}

\newcommand*{\INFNFE}{INFN, Sezione di Ferrara, 44100 Ferrara, Italy}
\newcommand*{\INFNFEindex}{21}
\affiliation{\INFNFE}
\newcommand*{\INFNFR}{INFN, Laboratori Nazionali di Frascati, 00044 Frascati, Italy}
\newcommand*{\INFNFRindex}{22}
\affiliation{\INFNFR}
\newcommand*{\INFNGE}{INFN, Sezione di Genova, 16146 Genova, Italy}
\newcommand*{\INFNGEindex}{23}
\affiliation{\INFNGE}
\newcommand*{\INFNRO}{INFN, Sezione di Roma Tor Vergata, 00133 Rome, Italy}
\newcommand*{\INFNROindex}{24}
\affiliation{\INFNRO}
\newcommand*{\INFNTUR}{INFN, Sezione di Torino, 10125 Torino, Italy}
\newcommand*{\INFNTURindex}{25}
\affiliation{\INFNTUR}
\newcommand*{\INFNCAT}{INFN, Sezione di Catania, 95123 Catania, Italy}
\newcommand*{\INFNCATindex}{26}
\affiliation{\INFNCAT}
\newcommand*{\INFNPAV}{INFN, Sezione di Pavia, 27100 Pavia, Italy}
\newcommand*{\INFNPAVindex}{27}
\affiliation{\INFNPAV}
\newcommand*{\ORSAY}{Universit'{e} Paris-Saclay, CNRS/IN2P3, IJCLab, 91405 Orsay, France}
\newcommand*{\ORSAYindex}{28}
\affiliation{\ORSAY}
\newcommand*{\Juelich}{Institute fur Kernphysik (Juelich), Juelich, Germany}
\newcommand*{\Juelichindex}{29}
\affiliation{\Juelich}
\newcommand*{\JMU}{James Madison University, Harrisonburg, Virginia 22807}
\newcommand*{\JMUindex}{30}
\affiliation{\JMU}
\newcommand*{\KNU}{Kyungpook National University, Daegu 41566, Republic of Korea}
\newcommand*{\KNUindex}{31}
\affiliation{\KNU}
\newcommand*{\LAMAR}{Lamar University, 4400 MLK Blvd, PO Box 10046, Beaumont, Texas 77710}
\newcommand*{\LAMARindex}{32}
\affiliation{\LAMAR}
\newcommand*{\MISS}{Mississippi State University, Mississippi State, MS 39762-5167}
\newcommand*{\MISSindex}{34}
\affiliation{\MISS}
\newcommand*{\ITEP}{National Research Centre Kurchatov Institute - ITEP, Moscow, 117259, Russia}
\newcommand*{\ITEPindex}{35}
\affiliation{\ITEP}
\newcommand*{\UNH}{University of New Hampshire, Durham, New Hampshire 03824-3568}
\newcommand*{\UNHindex}{36}
\affiliation{\UNH}
\newcommand*{\NMSU}{New Mexico State University, PO Box 30001, Las Cruces, NM 88003, USA}
\newcommand*{\NMSUindex}{37}
\affiliation{\NMSU}
\newcommand*{\NSU}{Norfolk State University, Norfolk, Virginia 23504}
\newcommand*{\NSUindex}{38}
\affiliation{\NSU}
\newcommand*{\OHIOU}{Ohio University, Athens, Ohio  45701}
\newcommand*{\OHIOUindex}{39}
\affiliation{\OHIOU}
\newcommand*{\JLUGiessen}{II Physikalisches Institut der Universitaet Giessen, 35392 Giessen, Germany}
\newcommand*{\JLUGiessenindex}{40}
\affiliation{\JLUGiessen}
\newcommand*{\RPI}{Rensselaer Polytechnic Institute, Troy, New York 12180-3590}
\newcommand*{\RPIindex}{41}
\affiliation{\RPI}
\newcommand*{\URICH}{University of Richmond, Richmond, Virginia 23173}
\newcommand*{\URICHindex}{42}
\affiliation{\URICH}
\newcommand*{\ROMAII}{Universita' di Roma Tor Vergata, 00133 Rome Italy}
\newcommand*{\ROMAIIindex}{43}
\affiliation{\ROMAII}
\newcommand*{\MSU}{Skobeltsyn Institute of Nuclear Physics, Lomonosov Moscow State University, 119234 Moscow, Russia}
\newcommand*{\MSUindex}{44}
\affiliation{\MSU}
\newcommand*{\SCAROLINA}{University of South Carolina, Columbia, South Carolina 29208}
\newcommand*{\SCAROLINAindex}{45}
\affiliation{\SCAROLINA}
\newcommand*{\TEMPLE}{Temple University,  Philadelphia, PA 19122 }
\newcommand*{\TEMPLEindex}{46}
\affiliation{\TEMPLE}
\newcommand*{\UTFSM}{Universidad T\'{e}cnica Federico Santa Mar\'{i}a, Casilla 110-V Valpara\'{i}so, Chile}
\newcommand*{\UTFSMindex}{47}
\affiliation{\UTFSM}
\newcommand*{\INSUBRIA}{Universit\`{a} degli Studi dell'Insubria, 22100 Como, Italy}
\newcommand*{\INSUBRIAindex}{48}
\affiliation{\INSUBRIA}
\newcommand*{\BRESCIA}{Universit`{a} degli Studi di Brescia, 25123 Brescia, Italy}
\newcommand*{\BRESCIAindex}{49}
\affiliation{\BRESCIA}
\newcommand*{\MESSU}{Universit`{a} degli Studi di Messina, 98166 Messina, Italy}
\newcommand*{\MESSUindex}{50}
\affiliation{\MESSU}
\newcommand*{\GLASGOW}{University of Glasgow, Glasgow G12 8QQ, United Kingdom}
\newcommand*{\GLASGOWindex}{51}
\affiliation{\GLASGOW}
\newcommand*{\YORK}{University of York, York YO10 5DD, United Kingdom}
\newcommand*{\YORKindex}{52}
\affiliation{\YORK}
\newcommand*{\VIRGINIA}{University of Virginia, Charlottesville, Virginia 22901}
\newcommand*{\VIRGINIAindex}{53}
\affiliation{\VIRGINIA}
\newcommand*{\WM}{College of William and Mary, Williamsburg, Virginia 23187-8795}
\newcommand*{\WMindex}{54}
\affiliation{\WM}
\newcommand*{\YEREVAN}{Yerevan Physics Institute, 375036 Yerevan, Armenia}
\newcommand*{\YEREVANindex}{55}
\affiliation{\YEREVAN}
\newcommand*{\TUDa}{Technische Universität Darmstadt, Fachbereich Physik, Darmstadt, Germany}
\newcommand*{\TUDaindex}{56}
\affiliation{\TUDa}

\newcommand*{\NOWISU}{Idaho State University, Pocatello, Idaho 83209}
\newcommand*{\NOWBRESCIA}{Universit`{a} degli Studi di Brescia, 25123 Brescia, Italy}

\author{I. Korover}
\thanks{Equal Contribution}
\affiliation{\MIT}
\author{A.W. Denniston}
\thanks{Equal Contribution}
\affiliation{\MIT}
\author{ A. Kiral}
\affiliation{\MIT}
\author{A. Schmidt}
\affiliation{\GWU}
\author{A. Lovato}
\affiliation{\ANL}
\author{N. Rocco}
\affiliation{\FNAL}
\author{A. Nikolakopoulos}
\affiliation{\FNAL}
\author{L.B. Weinstein}
\affiliation{\ODU}
\author{E. Piasetzky}
\affiliation{\TAU}
\author{O. Hen}
\email[Contact Author \ ]{hen@mit.edu}
\affiliation{\MIT}
% ...

\collaboration{The CLAS Collaboration}
\noaffiliation

\date{\today}

% Make the title.

\begin{abstract}
  We report the first measurement of $x_B$-scaling in $(e,e'p)$
  cross-section ratios off nuclei relative to deuterium at large
  missing-momentum of $350 \leq p_{miss} \leq 600$ MeV/c.  The
  observed scaling extends over a kinematic range of
  $0.7 \leq x_B \leq 1.8$, which is significantly wider than
  $1.4 \leq x_B \leq 1.8$ previously observed for inclusive $(e,e')$
  cross-section ratios.  The 
  $x_B$-integrated cross-section ratios become constant (i.e., scale)
  beginning at $p_{miss}\approx k_F$, the
  nuclear Fermi momentum.  Comparing with theoretical
  calculations we find good agreement with Generalized Contact
  Formalism calculations for high missing-momentum ($> 375$ MeV/c),
  suggesting the observed scaling results from interacting with
  nucleons in short-range correlated (SRC) pairs.  For low
  missing-momenta, mean-field calculations show good
  agreement with the data for  $p_{miss}\le k_F$, and suggest that contributions
  to the measured cross-section ratios from scattering off single,
  un-correlated, nucleons are non-negligible up to $p_{miss}\approx 350$ MeV/c.
  Therefore,  SRCs become dominant in nuclei at $p_{miss}\approx 350$
  MeV/c, well  above the
  nuclear Fermi Surface of $k_F \approx  250$ MeV/c.
\end{abstract}

\maketitle

Atomic nuclei are complex quantum-mechanical systems that account for most of the visible mass in the universe.
The complexity of the strong nuclear interaction makes it difficult to
use scattering reactions to experimentally probe the
detailed distributions of nucleons inside nuclei.  Experimental
nuclear physicists thus work together with theorists to find
measurable reactions that are sensitive to particular aspects of
nuclear dynamics.

By using high-energy electron beams to knock out nucleons from nuclei
in nearly elastic kinematics, one can learn about the behavior of
single nucleons in the nucleus~\cite{kelly96}.
This behavior can be generally explained by nucleons moving in nuclear
shell-model states (e.g. $s$-, $p$-, $d$-, ... shells) where the
typical nucleon momenta in each shell is smaller than the nuclear
Fermi momentum ($k_F$).  Full shell-model calculations improve on this
by introducing effective long-ranged correlations between the
nucleons~\cite{Dickhoff:2004xx}, which leads to the formation of a
nuclear Fermi Surface.

While these models can successfully describe the long-range structure
of nuclei, they do not describe the explicit high-resolution effects
of short-range correlated (SRC) nucleon pairs. Within a
high-resolution picture of nuclear dynamics, SRC pairs arise when two
nucleons get so close to each other that the short-range nuclear
interaction between them is much stronger than the effective
long-ranged nuclear mean field due to their interactions with all the
other nucleons in the nucleus~\cite{Frankfurt88,Atti:2015eda}.

SRCs have been clearly identified in data using large
momentum-transfer nucleon knockout
reactions~\cite{tang03,subedi08,Hen:2014nza,Patsyuk:2021jea,Hen:2016kwk}. They
are characterized by a high (greater than $k_F$) relative momentum
between the nucleons of the pair and are predominantly proton-neutron
pairs formed due to the action of the spin-dependent tensor part of
the strong nuclear interaction~\cite{schiavilla07, sargsian05,
  Alvioli:2007zz, neff15}.
They thus deplete the occupancy of single-particle shell-model states (below $k_F$) and populate high-momentum states~\cite{Frankfurt88,egiyan06,Atti:2015eda,Hen:2016kwk,Paschalis:2018zkx,Ryckebusch:2019oya}
While shell structures vary among nuclei, SRCs are a universal
phenomenon, i.e., they are similar in all
nuclei~\cite{CiofidegliAtti:1995qe,Atti:2015eda,ryckebusch15,Cruz-Torres:2019fum},
varying primarily in their magnitude.

A complete high-resolution microscopic description of atomic nuclei should thus have the nucleus-dependent mean field and long-ranged nuclear shell model parts as well as explicit nucleus-independent effects of SRC pairs.

Here we study the onset of SRC dominance in semi-inclusive high-energy
electron scattering reactions, where we detect the knocked-out proton
in addition to the scattered electron, $(e,e'p)$.  For the first time
in $(e,e'p)$ reactions, we observed scaling in the cross-section
ratios of nuclei from carbon to lead relative to deuterium over a broad
range in the Bjorken scaling variable, $x_B$.  This scaling
substantially extends the kinematical range where SRCs can be
identified and studied, as compared with previous inclusive $(e,e')$
measurements.  Thereby, they provide direct experimental evidence for
the dominance SRCs in the scattering response at high missing momenta,
and allow quantifying the onset of this dominance.

Our experiment ran at the Thomas Jefferson National Accelerator
Facility. It used a 5.01~GeV electron beam incident on a target system
consisting of a deuterium cell followed by an interchangeable solid
foil of carbon (C), aluminum (Al), iron (Fe), or lead
(Pb)~\cite{Hakobyan:2008zz}. Scattered electrons and knocked-out
protons were identified and measured using the CEBAF Large Acceptance
Spectrometer (CLAS) ~\cite{Mecking:2003zu} (see supplementary
materials for details).

In high-energy scattering, the electron transfers a single virtual
photon to the nucleus with momentum $\vec{q}$ and energy $\omega$.  In
the high-resolution quasielastic (QE) reaction picture, the virtual
photon is absorbed by a single nucleon, which gets knocked-out of the
nucleus with momentum $\vec{p}_p$.  By measuring both the scattered
electron and knocked-out proton, i.e. the $(e,e'p)$ reaction, we can
determine the missing momentum $\vec{p}_{miss} = \vec{p}_p - \vec{q}.$
The reaction is further characterized by the four-momentum transfer
$Q^2 = \vec q\thinspace^2 - \omega^2$ and Bjorken scaling variable
$x_B = Q^2/2m\omega$ where $m$ is the nucleon mass.

If the knocked-out nucleon does not re-interact as it leaves the
nucleus, then $\vec{p}_{miss}$ is equal to the initial momentum of
that nucleon.  Thus we expect the reaction to be sensitive to
mean-field nucleons at low-${p}_{miss}$ and to SRCs at
high-${p}_{miss}$~\cite{duer18}.  In the SRC dominated region, the
cross section ratio for any two nuclei should be constant (i.e.,
independent of $p_{miss}$) and equal to the relative number of
high-momentum nucleons in the two
nuclei~\cite{frankfurt93,egiyan02,egiyan06,fomin12,Schmookler:2019nvf,Atti:2015eda,Hen:2016kwk}.
Thus, by measuring the $(e,e'p)$ cross section ratio for nuclei
relative to deuterium for different minimum ${p}_{miss}$, we can
establish the onset of scaling that corresponds to SRC pair dominance
in the nuclear momentum distribution.

\begin{figure}
\centering
\includegraphics[width=\columnwidth]{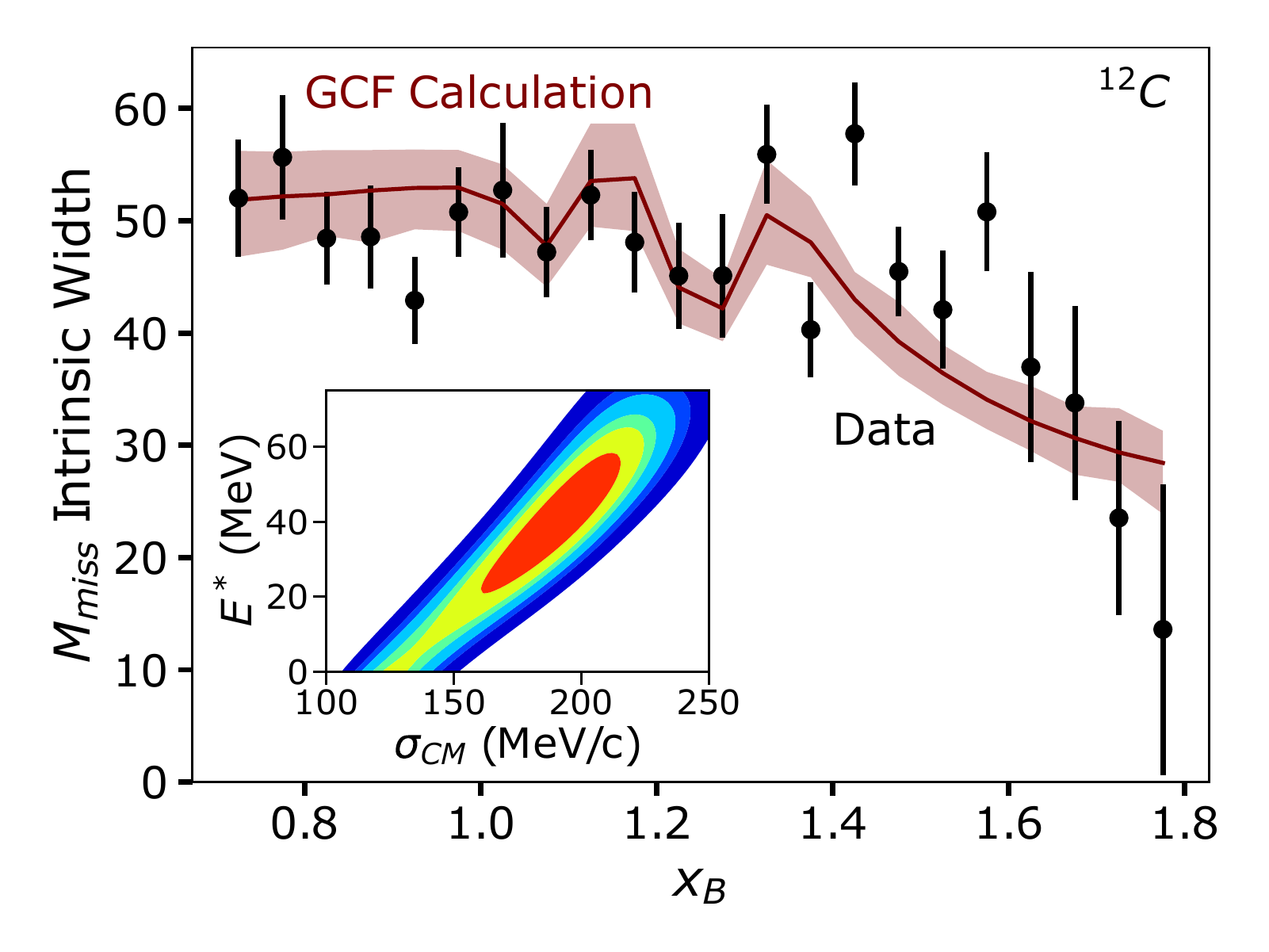}
\let\nobreakspace\relax
\caption{\label{fig:cm_width}
{\normalfont  The intrinsic width of the $^{12}$C missing mass ($M_{miss}$) distribution, plotted vs $x_B$. Black points show the data.  The red curve and uncertainty band shows an SRC based Generalized Contact Formalism (GCF) calculation~\cite{Cruz-Torres:2019fum,Pybus:2020itv}. The two main model parameters of the calculation, namely SRC pair CM momentum distribution width $\sigma_{CM}$ and the residual $A-2$ system excitation energy $E^*$, are fit to the data. Data error bars and calculation error band show the total uncertainty (statistical + systematical) at the $1\sigma$ or $68\%$ confidence level. Inset: The resulting confidence intervals of the correlation between the fitted values of $\sigma_{CM}$ and $E^*$. The inner region (red) shows the $1\sigma$ ($68\%$) confidence region with each region increasing the confidence by an additional $1\sigma$. The observed agreement between the data and GCF calculation, and the agreement of the fitted model parameters with previous extractions, show the measured $(e,e'p)$ events are consistent with resulting from the hard breakup of SRC pairs. 
}}
\end{figure}

To study this, we measured the $(e,e'p)$ reaction in conditions
sensitive to the knockout of protons from SRC pairs. We required
$Q^2 \ge 1.5$~(GeV/c)$^2$ and
$350 \le p_{miss}\le 600$ MeV/c to ensure a high-resolution reaction
that can resolve single nucleons in SRC pairs. We further required
that the proton be emitted within $25^\circ$ of the momentum transfer,
to ensure that the measured proton was the nucleon that absorbed the
virtual photon~\cite{hen14,schmidt20}.

We suppressed inelastic (non-QE) scattering events by requiring
$M_{miss}$, the missing mass for $(e,e'p)$ scattering from a
two-nucleon pair at rest, to be smaller than the nucleon mass ($m$)
plus pion mass ($m_\pi$), $0.8 \le M_{miss}\le m + m_\pi = 1.08$
GeV/c$^2$.  In non-QE reactions the momentum transferred to
undetected particles (e.g., pions) shifts the direction of
$\vec p_{miss}$. Therefore such events will have a larger
$\theta_{\vec{p}_{miss},\vec{q}}$, the angle between
$\vec p_{miss}$ and $\vec q$.  We
thus further suppressed the small non-QE tail below $M_{miss} = 1.08$
GeV/c$^2$ by observing that the measured
$\theta_{\vec{p}_{miss},\vec{q}}$ distribution has two maxima,
corresponding to QE and non-QE scattering, and selecting events in the
$\theta_{\vec{p}_{miss},\vec{q}}$ QE peak.  See Figs. S1 and S2 and
supplementary materials for details.

We tested our identification of scattering from protons in SRC
pairs by comparing the measured width of the $M_{miss}$ peak with that
calculated using the Generalized Contact Formalism (GCF) model for SRC
pairs (see Fig.~\ref{fig:cm_width})~\cite{Weiss:2015mba,Weiss:2016obx,Cruz-Torres:2019fum,
  schmidt20,Pybus:2020itv,Patsyuk:2021jea}.  This width depends on the
CLAS resolution and on the SRC pair center of mass (CM) motion.  We
corrected for the effects of the CLAS resolution by subtracting the
deuterium $M_{miss}$ peak width from that of $^{12}$C to get the
intrinsic width:
$\sigma^{^{12}C}_{int} = \sqrt{ (\sigma^{^{12}C}_{exp})^2 -
  (\sigma^{d}_{exp})^2 }$.

The measured $x_B$ dependence of $\sigma^{^{12}C}_{int}$ agrees well
with a GCF calculation that assumes electron scattering from nucleons
in SRC pairs with a realistic Gaussian CM momentum
distribution~\cite{Cohen:2018gzh}, as was done in
Refs.~\cite{schmidt20,Pybus:2020itv,Patsyuk:2021jea}.  The calculation
accounts for the CLAS detector acceptance and resolution and our event
selection cuts. The width of the CM momentum distributions,
$\sigma_{CM}$, and the excitation energy of the residual nuclear
system after the SRC breakup, $E^{*}$, were the only two free
parameters used in the calculation and were determined from a fit of
the calculation to the data (see inset of Fig.~\ref{fig:cm_width}).
For $\sigma_{CM}$ the fitted values of $160 - 210$ MeV/c ($125 - 220$)
at $68\%$ ($90\%$) confidence agree well with previous direct
measurements~\cite{Cohen:2018gzh,Patsyuk:2021jea}.
For $E^{*}$, while not previously measured, the fitted values of
$20 - 55$ MeV ($0 - 70$ MeV) at $68\%$ ($90\%$) confidence are
consistent with those used by previous analyses~\cite{schmidt20}.  The
sensible values of the resulting fit parameters and the agreement
between the $x_B$ dependence of the GCF calculation and the data
further support our interpretation of the data as dominated by
scattering off SRC pairs.

\begin{figure}[t]
\includegraphics[width=\columnwidth]{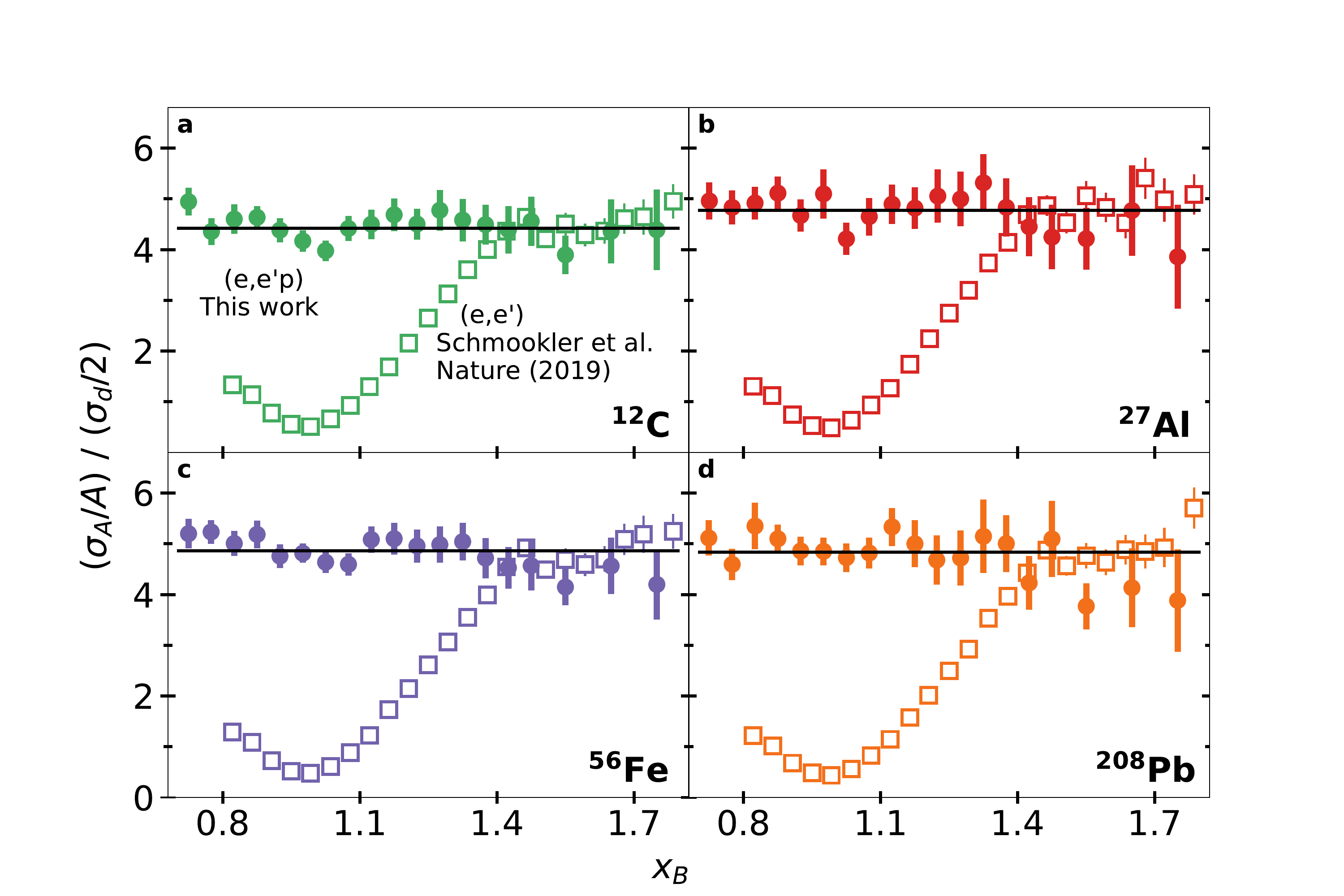}
\caption{\label{fig:all_nuclei} {\normalfont Measured $(e,e'p)$ per
    nucleon cross-section ratios for $350 \le p_{miss} \le 600$~MeV/c
    for carbon, aluminum, iron and lead relative to deuterium as
    function of $x_B$.  Open squares are the 
    inclusive $(e,e')$ per nucleon cross section ratios of
    Ref.~\cite{Schmookler:2019nvf}. The horizontal lines show the
    average $(e,e')$ cross section ratio for
    $1.45 \le x_B \le 1.8$~\cite{Schmookler:2019nvf}.  Error bars show
    the data uncertainty (statistical plus point-to-point
    systematical) at the $1\sigma$ or $68\%$ confidence level.
    Overall $(e,e'p)$ systematic uncertainties of 10\% (C) to 15\%
    (Pb) are not shown.  }}
\end{figure}

Using the selected event samples, we extracted the $(e,e'p)$ cross
section ratios for scattering off the solid targets relative to
deuterium.  We first divided the ratio of the measured numbers of
events for a given target to deuterium with the ratio of the
experimentally determined integrated luminosities to obtain the
normalized-yield ratios.  We then determined the cross section ratios
by correcting the normalized-yield ratios for attenuation of the
outgoing protons as they traverse the different
nuclei~\cite{Duer:2018sjb}, electron radiation effects, and the small
difference in the CLAS acceptance for detecting particles emitted from
the deuterium and the solid targets. At the large
$Q^2$ of this measurement, the attenuation correction is less
sensitive to the initial nucleon momenta and therefore both mean-field
and SRC breakup reactions have the same
attenuation~\cite{Duer:2018sjb}. Acceptance effects were calculated
using the CLAS detector simulation~\cite{CLAS-Geant} and an electron
scattering reaction event generator based on the GCF as applied in
previous studies~\cite{schmidt20,Pybus:2020itv} (see supplementary
materials for details).

Figure~\ref{fig:all_nuclei} shows the per nucleon $(e,e'p)$ cross section ratios for $350 \leq {p}_{miss} \leq 600$~MeV/c for carbon, aluminum, iron, and lead relative to deuterium as a function of $x_B$. 
The $(e,e'p)$ ratios scale (i.e., are constant) for all four nuclei
over the entire measured $x_B$ range.  This implies that the reaction
is probing similar nuclear configurations in the measured nuclei and
in deuterium.  As the deuteron is a simple proton-neutron correlated
two-body system, we interpret this high missing-momentum scaling as
observation of deuteron-like proton-neutron SRC pairs in nuclei.
The cross-section ratio is thus a measure of their relative abundance.

This interpretation is supported by the consistency between our
measured $(e,e'p)$ cross section ratios and previously measured
inclusive $(e,e')$ scattering cross section ratios at similar $Q^2$
and at $x_B\ge 1.5$~\cite{frankfurt93,egiyan02,egiyan06,fomin12,Schmookler:2019nvf} (see open symbols in Fig.~\ref{fig:all_nuclei}).
As the inclusive scaling onset at $x_B \approx 1.5$ has been attributed  to scattering off nucleons with momenta greater than $\sim 275$ MeV/c~\cite{egiyan06,Weiss:2021zyb}, it is also associated with scattering off nucleons in deuteron-like proton-neutron SRC pairs, formed by the strong tensor interaction~\cite{frankfurt93,Schmookler:2019nvf} (see supplementary materials for details).
Proton detection extends the cross-section ratio plateau down to $x_B=0.7$, providing a new experimental tool to study the transition to SRC dominance in nuclei over a broad range in $x_B$.

\begin{figure} [t]
\centering
\includegraphics[width=\columnwidth]{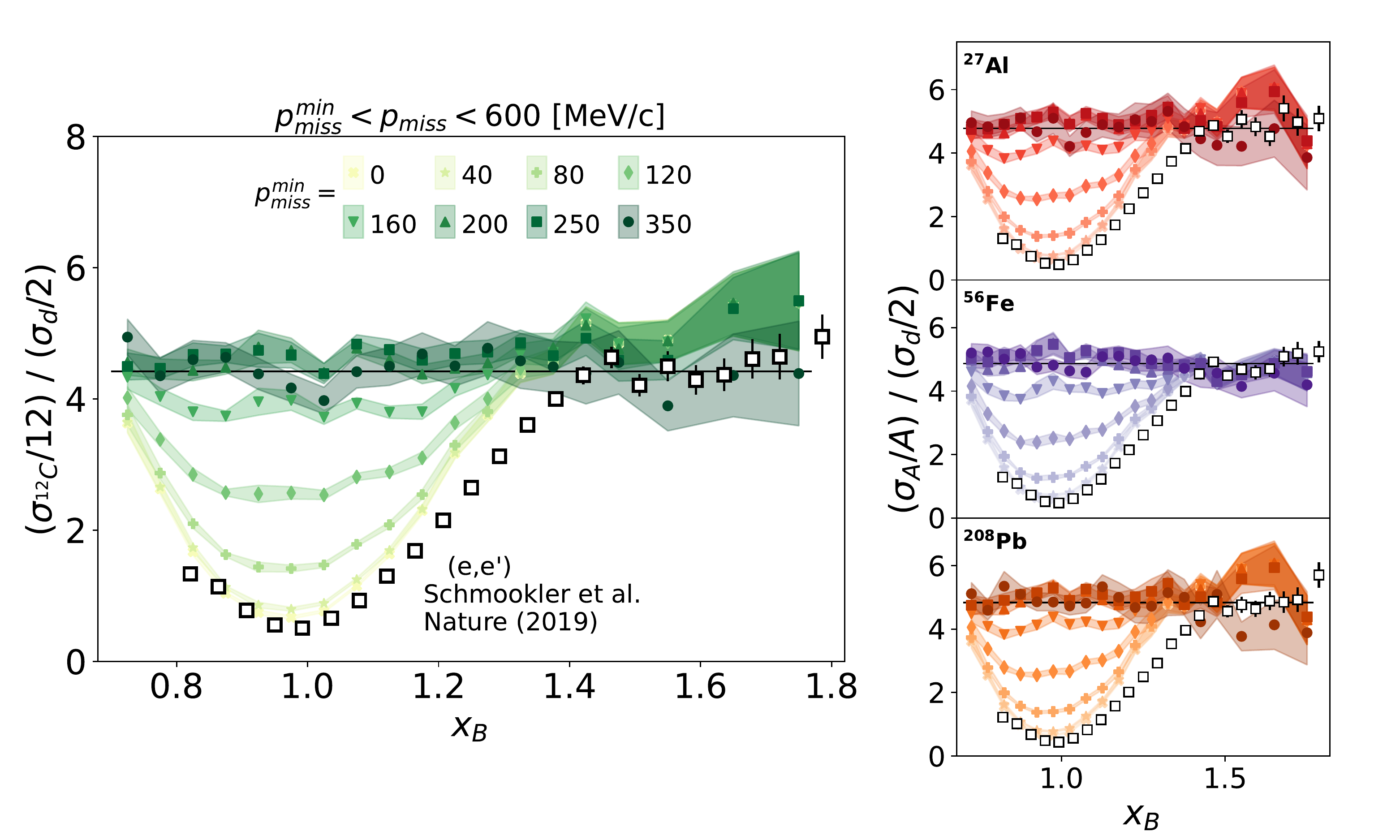}
\caption{\label{fig:C_xB_Pm} {\normalfont The per-nucleon
    cross-section ratios for carbon (left), aluminum, iron and lead
    (right) to deuterium as a function of $x_B$. Full symbols with
    different colors indicate different lower limits of the $p_{miss}$
    integration. The upper $p_{miss}$ limit is fixed at
    $600$~MeV/c. The colored bands mark the statistical plus
    point-to-point systematical uncertainty of the data at the
    $\pm1\sigma$ or $68\%$ confidence level. Overall systematic
    uncertainties of 10--15\% are not shown.  Open black squares show
    the inclusive $(e,e')$ per nucleon cross-section ratios of
    Ref.~\cite{Schmookler:2019nvf}. The horizontal line shows the
    average $(e,e')$ cross section ratio for
    $1.45 \le x_B \le 1.8$~\cite{Schmookler:2019nvf}.  }}
\end{figure}

We next examined how this scaling depends on the minimum $p_{miss}$
range of the data.
Figure~\ref{fig:C_xB_Pm} shows the per nucleon $(e,e'p)$ cross section ratios for the measured nuclei relative to deuterium as a function of $x_B$ for different minimum $p_{miss}$.  
For all nuclei, the curve for $p_{miss}^{min}=0$ are similar to the inclusive data of Schmookler et al.~\cite{Schmookler:2019nvf}, with  a plateau for $x_B \ge 1.5$ and a minimum at $x_B \approx 1$.  
As $p_{miss}^{min}$ increases, this minimum fills in.  For
$p_{miss}^{min} \gtrsim 200 - 250$~MeV/c, it is completely filled in
and the $(e,e'p)$ cross section ratio scales over the full measured
$x_B$ range of $0.7$ to $1.8$.  This indicates that short-range
interactions become dominant at around
$k_F\approx 220 - 260$~MeV/c~\cite{moniz71}, as expected.

\begin{figure}
\centering
\includegraphics[width=\columnwidth]{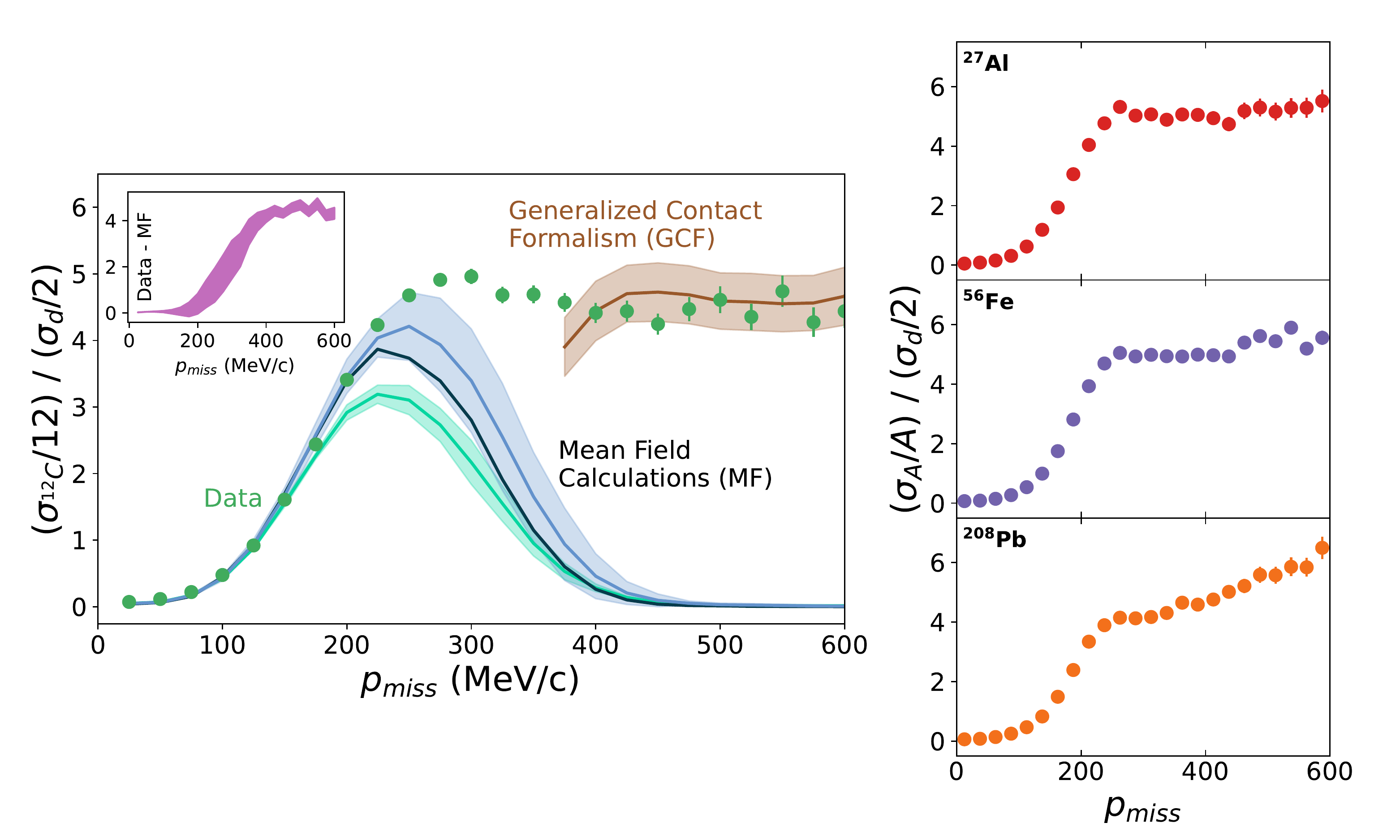}
\caption{\label{fig:C_d_PmRatio} {\normalfont The per nucleon
    $(e,e'p)$ cross-section ratios for different nuclei to deuterium
    as a function of $p_{miss}$, integrated over $0.7\le x_B\le 1.8$.
    Filled circles show the data.  The teal, black, and azure lines
    show the calculated cross sections obtained using QMC (teal), IPSM
    (black), and Skryme (azure) based one-body mean-field models for
    nucleon distributions in carbon. The brown line shows a GCF
    calculation for SRC nucleons in carbon. Data error bars and the
    widths of the bands around the calculation lines show their
    uncertainties (statistical plus point-to-point systematical) at
    the $1\sigma$ or $68\%$ confidence level.  Overall data systematic
    data uncertainties of 10--15\% are not shown.  The insert shows the
    result of subtracting the mean-field calculations from the carbon
    data.  }}
\end{figure}

To better quantify this transition, Figure~\ref{fig:C_d_PmRatio} shows
the $p_{miss}$ dependence of the $(e,e'p)$ cross section ratio for the
different nuclei relative to deuterium, integrated over the scaling
regions of $0.7\le x_B\le 1.8$.  The measured cross section ratio for
carbon ($k_F\approx220$~MeV/c), aluminum ($k_F\approx235$~MeV/c), and
iron ($k_F\approx260$~MeV/c) all become flat starting around the Fermi
momentum at $p_{miss}\approx 250$.  The lead ratio shows a similar
transition but does not fully plateau, possibly owing to its much
larger neutron-to-proton ratio or to increased final state interactions
due to its larger size.

We thus conclude that the data indicate the existence of a clear
transition in the nuclear response around the nuclear Fermi momentum,
resulted by the onset of the SRC dominance at high-momenta.

Focusing on the carbon nucleus, where theoretical calculations are
readily available, we find that the high-$p_{miss}$ data are in
excellent agreement with an asymptotic GCF calculation of the
cross-section ratio (brown band in Fig.~\ref{fig:C_d_PmRatio}, left
panel).  The calculation was done using a factorized plane-wave
impulse approximation (PWIA) for the scattering reaction, with SRC-pair spectral functions calculated using the GCF~\cite{Weiss:2018tbu}
and transparency and single-charge exchange corrections as done in
Ref.~\cite{Duer:2018sxh, schmidt20, Pybus:2020itv} (see supplementary
materials for details).  The SRC-pair abundance parameters used by
the GCF calculation were all previously determined by $ab-initio$
many-body theoretical calculations~\cite{Cruz-Torres:2019fum},
offering additional support to our identification of QE scattering
events and the dominance of interacting with nucleons in SRC pairs at
high-$p_{miss}$.

Lastly, we estimate the possible contribution of single-nucleon
(one-body) states to the measured cross-section ratio around $k_F$, to
assess their impact on the observed SRC scaling onset.
We examined three calculations using different single-nucleon spectral
functions: (1) Independent particle shell-model with Woods-Saxon
potential~\cite{oneill95,Makins:1994mm}, (2) Skryme model using 5
different functionals~\cite{Waroquier:1979zz,Chabanat:1997un} and (3)
new Quantum Monte-Carlo (QMC) many-body calculations of the overlap
between the $^{12}$C and $^{11}$B+proton wave functions (see
supplementary materials for details). For the latter, we added the
contributions from both the ground state and a range of $^{11}$B
excited states to include a wide range of mean-field, single-nucleon
states. These models each assume very different underlying single-body
nuclear dynamics, and thus the spread in their results offers a
general measure for the model dependence of the single-body mean-field
contribution.

The resulting calculated one-body mean-field contributions to the C/d
$(e,e'p)$ cross section ratio are shown in
Figure~\ref{fig:C_d_PmRatio}, left panel. The calculations were done
using the same factorized PWIA scheme as for the GCF calculation, only
using mean-field spectral functions. The IPSM and Skryme calculations
are re-normalized (quenched) to agree with our low-$p_{miss}$
($\leq 150$ MeV/c) high-$Q^2$ data. This effectively accounts for
their lack of single-nucleon strength lost to long- and short-ranged
correlations and/or few-body reaction operators that can compensate
for it~\cite{Tropiano:2021qgf}. In contrast, the QMC calculation
extracts the underlying single-nucleon states from the fully
correlated high-resolution wave function. It thus has fewer than six
protons in its mean-field orbitals and does not require additional
quenching..  The agreement of the QMC calculation with the
low-$p_{miss}$ data thus confirms the completeness of the calculation.

The different single-nucleon calculations are similar and all show  the existence of
residual single nucleon contributions above $k_F$.  
We subtracted the calculated one-body mean-field contribution from
the measured cross-section ratio(see the inset in
Fig.~\ref{fig:C_d_PmRatio})left.
Accounting for these contributions can shift the scaling onset from the
purely experimental onset at $\sim k_F$ to a higher value of $\sim 350$
MeV/c. Such a shift would be consistent with the existence of a
nuclear Fermi-surface that accounts for long-range correlations above
$k_F$.

To conclude, the new nuclear scaling measurements presented herein
allow isolating interactions with SRC pairs in a
substantially-extended kinematical regime.  By examining the scaling
onset in missing momentum, we observe a universal transition in the
scattering response above the nuclear Fermi momentum.  Using model
dependent estimates for mean field contributions, we see an indication
for the onset of full SRC dominance above
$\sim 350$ MeV/c. Detailed theoretical calculations will be able to
use our data to fully quantify this mean-field to SRC transition
region and to obtain an effective high-resolution description of a
wide range of heavy nuclei.

\begin{acknowledgments}
We acknowledge the efforts of the staff of the Accelerator and Physics Divisions at Jefferson Lab that made this experiment possible. The analysis presented here was carried out as part of the Jefferson Lab Hall B data-mining project supported by the U.S. Department of Energy (DOE). The research was also supported by the National Science Foundation, the Israel Science Foundation, the Pazi Foundation, the Chilean Comisión Nacional de Investigación Científica y Tecnológica, the French Centre National de la Recherche Scientifique and Commissariat a l'Energie Atomique, the French–American Cultural Exchange, the Italian Istituto Nazionale di Fisica Nucleare, the National Research Foundation of Korea, and the UK Science and Technology Facilities Council. Jefferson Science Associates operates the Thomas Jefferson National Accelerator Facility for the DOE, Office of Science, Office of Nuclear Physics under contract DE-AC05-06OR23177.
\end{acknowledgments}

\bibliography{AoverD_ep_v13_arXiv.bbl}

\end{document}